\documentclass[10pt]{article}
\usepackage{opex2}
\usepackage{graphics,color,epsfig}
\begin{document}
\title{Inversionless gain in an optically-dense resonant Doppler-broadened medium}
\author{A. K. Popov and S. A. Myslivets}
\address{Institute for Physics, Russian Academy of Sciences,
Krasnoyarsk, 660036, Russia }
\email{popov@ksc.krasn.ru}
\smallskip
\author{Thomas F. George}
\address{Office of the Chancellor / Departments of Chemistry
and Physics \& Astronomy, University of Wisconsin-Stevens Point,
Stevens Point, WI 54481-3897, USA}
\email{tgeorge@uwsp.edu}

\begin{abstract}
Resonant nonlinear-optical interference processes in four-level Doppler-broadened
media are studied. Specific features of amplification and optical switching of
short-wavelength radiation in a strongly-absorbing resonant gas under coherent quantum
control with two longer wavelength radiations, are investigated. The major outcomes
are illustrated with virtual experiments aimed at inversionless short-wavelength
amplification, which also address deficiencies in this regard in recent experiments.
With numerical simulations related to the proposed experiment in optically-dense
sodium dimer vapor, we show optimal condition for optical switching and the expected
gain of the probe radiation, which is above the oscillation threshold.
\end{abstract}
% The commands,
\begin{OEReferences}
%scul1,koch,levi,man,scul2,rau,bull,har
\bibitem[$\dag$]{a} Ossipov Russian Folk Orchestra, condacted by Nikolai
Kalinin, ``Ochi Chornyje'' \copyright 1994 CD Ltd.
\bibitem[$\ddag$]{b}Thomas F. George, piano, Michael Kaupa, trumpet/fl\"ugelhorn,
``Close Your Eyes'' \copyright 1995 Hester Park,
\url{http://www.uwsp.edu/admin/chancellor/tgeorge/CDreview.htm}
\bibitem{scul1} M.O. Scully, ``Resolving conumdrums in lasing without
inversion via exact solutions to simple models'' Quantum Optics {\bf 6}, 203-215
(1994).
\bibitem{koch}O. Kocharovskaya and P.Mandel, ``Basic models of lasing  without inversion:
General form of amplification condition and problem  of self-consistency,'' $ibid$.,
217-230.
\bibitem{levi}B. G. Levi, ``Some benefits of quantum interference become transparent,''
Physics Today {\bf 45}, 17-19 (May 1992).
\bibitem{man}P. Mandel, ``Lasing without inversion: a
useful concept?'' Contemp. Phys. {\bf 34}, 235-246 (1993).
\bibitem{scul2}M. O. Scully and M. Fleischhauer, ``Laser without inversion,'' Science {\bf 263},
337-338 (1994).
\bibitem{rau}A. K. Popov and S. G. Rautian, "Atomic coherence and interference phenomena in resonant
nonlinear optical interactions" (invited paper), in \underline{Coherent Phenomena and
Amplification without Inversion}, A. V. Andreev, O. Kocharovskaya and P. Mandel, eds.,
Proc. SPIE  {\bf 2798}, 49-61 (1996), \url{http://xxx.lanl.gov/abs/quant-ph/0005114}.
\bibitem{bull}A.~K. Popov, "Inversionless
amplification and laser-induced transparency at discrete transitions and the
transitions to continuum" (review), Bull. Russ. Acad. Sci., Physics  {\bf 60},
927-945 (1996), \url{http://xxx.lanl.gov/abs/quant-ph/0005108}.
\bibitem{har}S.~E. Harris, ``Electromagnetically induced transparency,'' Physics Today {\bf 50}, 36-42 (July
1997).
%mer,hau,yam
\bibitem{mer} A. J. Merriam, S. J. Sharpe, H. Xia, D. Manuszak,
G. Y. Yin, and S. E. Harris, ``Efficient gas-phase generation of coherent vacuum
ultraviolet radiation,'' Opt. Lett. {\bf 24}, 625-627 (1999).
\bibitem{hau}S.~E. Harris and L. V. Hau, ``Nonlinear optics in low light-levels,'' Phys. Rev. Lett. {\bf 82}, 4611-4614
(1999).
\bibitem{yam}S.~E. Harris and Y. Yamamoto, ``Photon switching by quantum interference,''
Phys. Rev. Lett. {\bf 81}, 3611-3614 (1999.
\bibitem{QN} M. D. Lukin, A. V. Matsko, M. Fleischhauer, and M.O. Scully,
``Quantum noise and correlations in resonantly enhanced wave mixing based on atomic
coherence,'' Phys. Rev. Lett. {\bf 82}, 1847-1850 (1999).
\bibitem{HIN} U. Hinze, L.Meyer, B. N. Chichkov, E. Tiemann, and B. Wellegehausen,
``Continuous parametric amplification in a resonantly driven double-$\Lambda$
system,'' Opt. Commun. {\bf 166}, 127-132 (1999).
%Mys,akp
\bibitem{Mys} A. K. Popov and S. A. Myslivets, ``Resonant four-wave frequency mixing
in Doppler-broadened transitions,''  Quantum Electron. {\bf 27} 1004-1008 (1997),
\url{http://turpion.ioc.ac.ru/}.
\bibitem{akp}A.~K. Popov, ``Interference at quantum transitions:
lasing without inversion and resonant four-wave mixing in strong fields at
Doppler-broadened transitions,'' in \underline{Nonlinear Optics}, Sergei G. Rautian,
Igor M. Beterov and Natalia M. Rubtsova, eds., Proc. SPIE {\bf 3485}, 252-263 (1998),
\url{http://xxx.lanl.gov/abs/quant-ph/0005118}.
\bibitem{NIE} T. Ya. Popova, A. K. Popov, S.  G. Rautian, and R. I. Sokolovskii, ``Nonlinear
interference effects in emission, absorption, and generation spectra,'' JETP {\bf
30}, 466 (1970) [Translated from Zh. Eksp. Teor. Fiz. {\bf 57} 850, (1969)],
\url{http://xxx.lanl.gov/abs/quant-ph/0005094}.
%AWI1,AWI2,AWI3
\bibitem{AWI1} T. Ya. Popova, A. K. Popov, ``Effect of resonance radiative processes
on the amplification factor,'' Zhurn.Prikl. Spektrosk., {\bf 12}, No 6, 989, (1970)
[Translated in Engl.:\ Journ. Appl. Spectr {\bf 12}, No 6, 734, (1970),
\url{http://xxx.lanl.gov/abs/quant-ph/0005047}.
\bibitem{AWI2}T. Ya. Popova, A. K. Popov, ``Shape of
the amplification line corresponding to an adjacent transition in a strong field,''
Izv.Vysh. Uchebn. Zaved., Fizika No 11, 38, (1970) [Translated in Engl.:\ Soviet Phys.
Journ. {\bf 13}, No 11, 1435, (1970)], \url{http://xxx.lanl.gov/abs/quant-ph/0005049}.
\bibitem{AWI3}A. K. Popov, {\it Introduction in Nonlinear Spectroscopy\/}, Nauka, Novosibirsk,
1983, 274p. (in Russ).
\bibitem{BET} I. M. Beterov, ``Investigation on nonlinear resonant interaction of optical fields
in three-level gas laser,'' Cand. Sci. Dissertation, Institute of
Semiconductor Physics SD USSR AS, Novosibirsk, Dec. 1970.
\bibitem{TAR} A. K. Popov, S. A. Myslivets, E. Tiemann, B. Wellegehausen and
G. Tartakovsky, ``Quantum interference and Manley-Rowe relations at
resonant four-wave mixing in optically-thick Doppler-broadened
Media,'' JETP Lett. {\bf 69}, 912-16 (1999),
\url{http://ojps.aip.org/jetplo/}.
\end{OEReferences}
%\noindent
\section{Introduction}
Coherent quantum control (CQC) of optical processes have proven to be a
powerful tool to manipulate refraction, absorption, transparency, gain and conversion
of electromagnetic radiation (for a review, see \cite{scul1,koch,levi,man,scul2,rau,bull,har}).
 Among recent achievements
are the slowing down of the light group speed to a few m/s, highly-efficient frequency
conversion, squeezed quantum state light sources and optical switches for quantum
information processing \cite{mer,hau,yam,QN}. Much interest has been shown in the physics and
diverse practical schemes of amplification without inversion (AWI). This present
paper is aimed at studying nonlinear coherence and interference effects (NIE)
underlying an approach which accounts for the specific features of Doppler broadening
of coherently-driven quantum transitions and the inhomogeneous distribution of the
power saturated material parameters along the medium. Special emphasis is placed on
NIE processes, which would allow converting easily-achievable long-wavelength gain to
a higher- frequency interval. Another outcome is the realization of the optical
switching of the medium from opaque to amplifying via a transparent state by small
variation of the intensity or frequency of one of the coupled radiations. Important
features of both the signal and idle radiation produced with the aid of the proposed
scheme entail the suppression of quantum noise. We propose potential experiments
towards inversionless generation and optical switch based on CQC. The potential and
optimum conditions for the realization of such effects in an optically-dense
Doppler-broadened medium are explored.
%%%%%%%%%%%%%%%%%%%%%%%%%%%%%%%%%%%%%%%%%%%%%%%%%%%%%%%%%%%%%
\section{Resonance coherence and interference processes at
Doppler broadened transitions: propagation features and
parametric gain}
%\section{Theory}
%\begin{floatingfigure}{2.6cm}
\begin{figure}
\begin{center}
%\hspace*{-5mm}
\includegraphics[width=2.5cm]{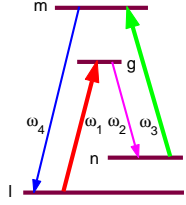}
\caption{\label{4lsh}Energy levels  and  coupled fields. }
\end{center}
\end{figure}
%\end{floatingfigure}
 A typical coupling schematic relevant to the topics under discussion and to
the experimental situation \cite{HIN} is depicted in Fig.~\ref{4lsh}. A weak probe
radiation $E_4$ at frequency $\omega_{4}$ propagates through an optically-dense
Doppler-broadened medium controlled with two longer-wavelength driving radiations
$E_1$ and $E_3$ at $\omega_{1}$ and $\omega_{3}$. All waves are co-propagating. Owing
to four-wave mixing (FWM), the probe wave generates radiation at
$\omega_{2}=\omega_{1}+\omega_{3}-\omega_{4}$ at the transition $gn$, where the
driving field $E_1$ easily produces a large Stokes gain. Alternatively, the enhanced
generated radiation $E_2$ contributes back to $E_4$ due to FWM, which dramatically
changes the propagation features of the probe field. Therefore, the problem under
consideration reduces to the solution of a set of the coupled equations  for the four
waves $(E_i/2)\, \exp [i(k_iz-\omega_i t)] + c. c.\, \, (i= 1...4)$, traveling in a
optically-thick medium:
\begin{eqnarray}
{dE_{4,2}(z)}/{dz}&= &i\sigma_{4,2} E_{4,2}+i\tilde\sigma_{4,2}
E_1E_3E_{2,4}^*, \label{e42}\\
{dE_{1,3}(z)}/{dz} &=& i\sigma_{1,3} E_{1,3}+
i\tilde\sigma_{1,3}E_4E_2E_{3,1}^*. \label{e13}
\end{eqnarray}
Here $\omega_4+\omega_2 =\omega_1+\omega_3$, $k_j$  are wave
numbers in a vacuum, $\sigma_j=-2\pi k_j \chi_j = \delta
k_j+i{\alpha_j}/{2}$, $\alpha_j$ and $\delta k_j$ are
intensity-dependent  absorption indices and dispersion parts of
the wave numbers, $\tilde\sigma_4=-2\pi k_4 \tilde\chi_4 $ (etc.)
is a FWM cross-coupling parameter,  and $\chi_j$ and
$\tilde\chi_j$ are the corresponding nonlinear susceptibilities
dressed by the driving fields. An amplification or  absorption of
any of the coupled radiations influences the propagation features
of the other ones. In the  the approximation that a change of the
driving radiations $ E _ {1,3} $ along the medium is neglected
(for example, at the expense of the saturation effects), the system
(\ref {e42})-(\ref {e13}) reduces to two coupled standard
equations of nonlinear optics for $ E _ {4} $ and $ E _ {2} $,
whereas the medium parameters are homogeneous along $ z $. If
$\Delta k=\delta k_1+\delta k_3-\delta k_2-\delta k_4=0$, the input
values $I_{4,2}\equiv |E_{4,2}|^2$ at $z=0$ are $I_{20}=0,
I_{40}\neq 0$ and the absorption (gain) rate substantially exceeds
that of the nonlinear optical conversion we obtain at the exit of
the medium of the length $L$:
\begin{eqnarray}
I_4/I_{40}\,=\,|\,\exp(-\alpha_4l/2)+
({\gamma^2}/{(2\beta)^2})\left[\exp(g_2L/2)-
\exp(-\alpha_4L/2)\right]|^2.\;\label{opa}
\end{eqnarray}
Alternatively, if $I_{40}=0, I_{20}\neq 0$,
\begin{eqnarray}
\eta_4 =\;I_4/I_{20}\,=\,({|\gamma_4|^2}/{(2\beta)^2})
\left|\exp(g_2L/2)\,- \exp(-\,\alpha_4 L/2)\right|^2. \label{fwm}
\end{eqnarray}
Here $\gamma^2=\gamma_2^*\gamma_4$,\
$\gamma_{4,2}=\chi_{4,2}E_1E_3$, $\beta = (\alpha_4-\alpha_2)/4$,
$(|\gamma^2|/\beta^2\ll 1)$, and $g_2\equiv -\alpha_2 $. From (\ref
{opa}) and (\ref {fwm}) it follows that at relatively small lengths the
FWM coupling may even increase the depletion rate of the probe
radiation, depending on the signs of $ Im \gamma _ {4,2} $ and $
Re \gamma _ {4,2}$. In order to achieve amplification, large
optical lengths $L$ and significant Stokes gain on the transition
$gn$ ($ \exp (g _ 2L/2) \gg | {(2\beta) ^ 2} / {\gamma ^ 2} | $),
as well as effective FWM both at $\omega_2$ and $\omega_4$, are
required.

The important feature of the considered far-from-degenerate interaction is that the
magnitude and sign of single-photon and multiphoton resonance detunings and,
consequently, of the amplitudes and phases of nonlinear polarizations, differ for
molecules at different velocities due to the difference in their frequency Doppler
shifts. Their interference determined by the interference of elementary quantum
pathways, and accounting for Maxwell's velocity distribution and saturation effects,
results in a nontrivial dependence of the material macroscopic parameters on the
intensities of the driving fields and on the frequency detunings from the
Doppler-broadened resonances.  The relevant solutions for elements of the density
matrix,  and consequently for the material parameters in the ED equations (\ref{e42})
and (\ref{e13}), are cumbersome. They are provided in \cite{Mys,akp}, where the solution
is found exactly with respect to $E_{1,3}$ and in the first approximation for
$E_{4,2}$. The various relaxation processes, Boltzmann excitation of the level $n$,
distribution over rotational sub-levels and dependence on the molecule velocity were
taken into account. This allows numerical averaging over the Maxwell velocity
distribution as well as further virtual experiments. These formulas are used here for
analysis of the dependence of the quantities $ \alpha _ {4,2}$ and $\gamma _ {4,2} $
on the intensities and frequencies of the fields and also to obtain a numerical
solution of the system (\ref {e42})--(\ref {e13}) which accounts for the
inhomogeneity of the coefficients and phase mismatch $ \Delta k $.

\section{Numerical simulations}
 The numerical simulations are done for the transitions $l - g - n - m - l$
(Fig. 1) attributed to those of sodium dimers [$Na_2$: $X'\Sigma_g^+(v"=0,J"=45)-
A'\Sigma^+_u(6,45)(\lambda_1= 655$ $ nm) - X^1\Sigma_g^+(14,45) (\lambda_2= 756$ $
nm) - B^1\Pi_u(5,45) (\lambda_3= 532$ $nm) - X'\Sigma_g^+(0,45) (\lambda_4= 480$
$nm)$] from the experiment \cite{HIN} and using experimental relaxation data. The
Doppler width of the transition (FWHM) at the wavelength $ \lambda _ 4 = 480 $ $ nm $
at a temperature of about 450$^\circ$ C is approximately equal to 1.7 GHz. Then the
Boltzmann population of level $ n $ is about 2\% of that of level $ l $. The driving
radiations are set at exact resonance with the corresponding transitions
($\Omega_1=\omega_1-\omega_{gl}=0, \Omega_3=\omega_3-\omega_{mn}=0$) and are
characterized by the coupling Rabi frequencies $G_1=E_1d_{lg}/2\hbar$ and
$G_3=E_3d_{nm}/2\hbar$. The resonant condition results in strong depletion of the
driving radiation $E_1$ along the medium. We scale the length of the medium $L$ to
the absorption length $L_4=1/\alpha_{40}$ at $\Omega_4 = \omega_4-\omega_{ml} = 0$,
with all driving fields turned off.

Figures 2(a-c) and the movie mov1a.mov  and mov1b.mov display changes in the spectral
dependencies of the velocity averaged material parameters with variation of the
driving fields. Figure 2(a) is generated for the input Rabi frequencies $G_{10} = 60$
MHz and $G_{30} = 20$ MHz. It displays substantial Autler-Townes splitting of the
Stokes lineshape with minimum gain in the resonance. Corresponding quantum
interference structures appear in the absorption and FWM coupling parameters, which
behave quite differently. As it is shown below, maximum gain $I_4(L)/I_{40}$ occurs
at $L /L_{40} = 15$, where the driving fields deplete down to $G_1 = 16$ MHz and $G_3
= 19$ MHz. This gives rise to a dramatic change in the spectral properties of
absorption (upper plot), of the Stokes gain (lower plot) and of the FWM coupling
parameters $\gamma_{4,2}$ (Fig. 2b). Due to the interplay of the quantum interference
processes and population cycling in the strong fields, absorption at the resonance
becomes even larger than in the unperturbed medium. Figure 2(c), computed for $G_1 =
60$ MHz and $G_3 = 80$, displays further feasibility to manipulate material
parameters with quantum coherence and interference processes.
\begin{figure}[!h]
\begin{center}
\includegraphics[width=.32\textwidth]{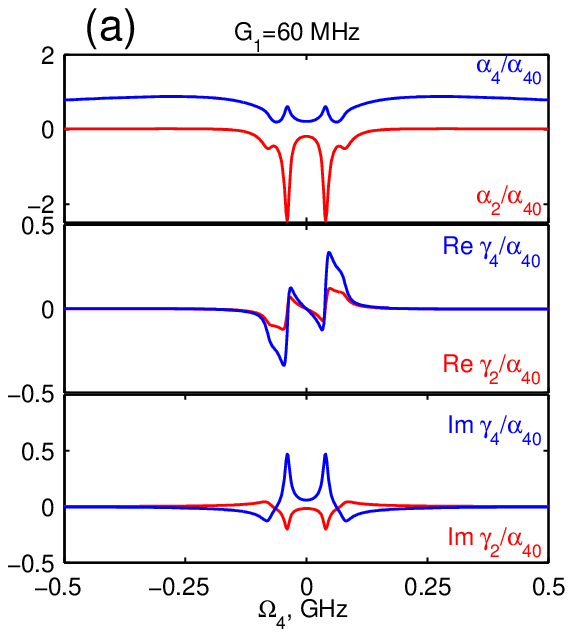}
\includegraphics[width=.32\textwidth]{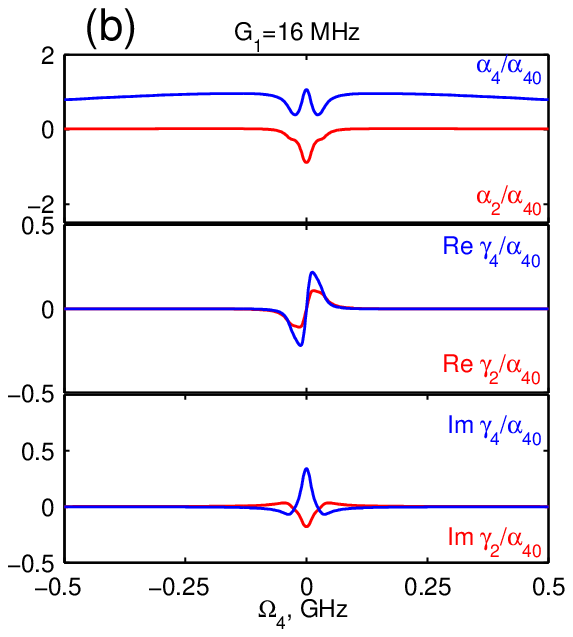}
\includegraphics[width=.32\textwidth]{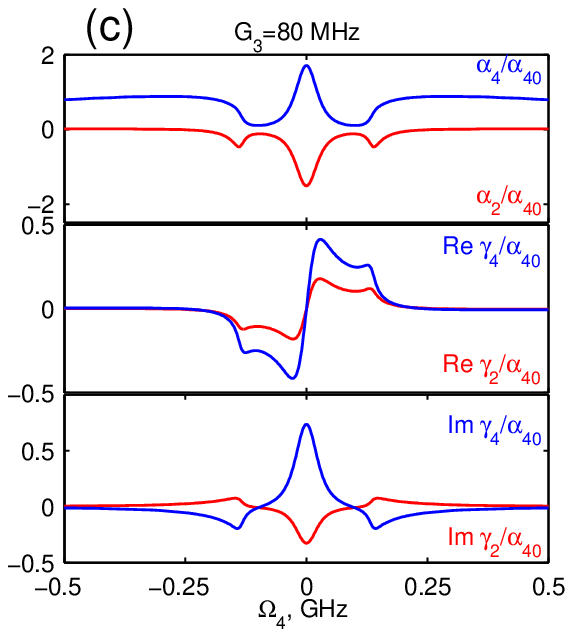}
\end{center}\vspace{-5mm}
\caption{Intensity-dependent absorption index for the signal radiation, Stokes
amplification index for the idle radiation and real and imaginary parts of the FWM
cross coupling parameters vs signal frequency resonance detuning. The driving fields
are set to the resonance. $(a)$ and $(b)$ -- mov1a.mov (2.02MB, audio$^\dag$)
($G_3=20$ MHz, $G_1$ varies); $(c)$ -- mov1b.mov (2.03MB, audio$^\dag$) ($G_1=60$
MHz, $G_3$ varies). The current magnitude of the variable Rabi frequency runs at the
top of the screen.}
\end{figure}
%\vspace{5mm}
\begin{figure}[!h]
\begin{center}
\includegraphics[width=.32\textwidth]{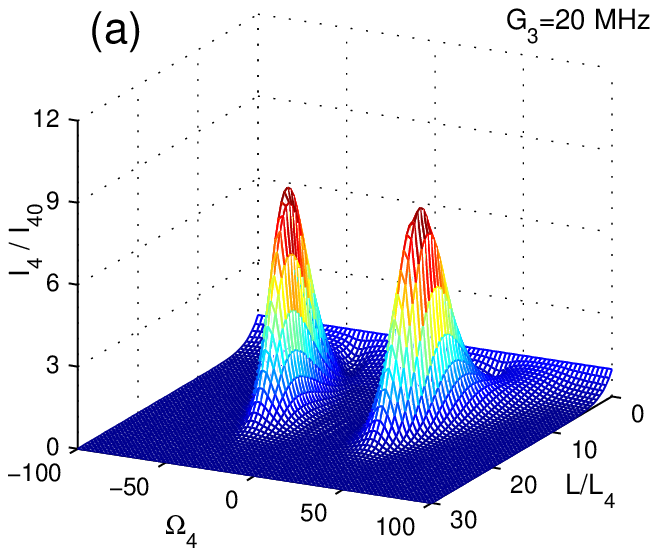}
\includegraphics[width=.32\textwidth]{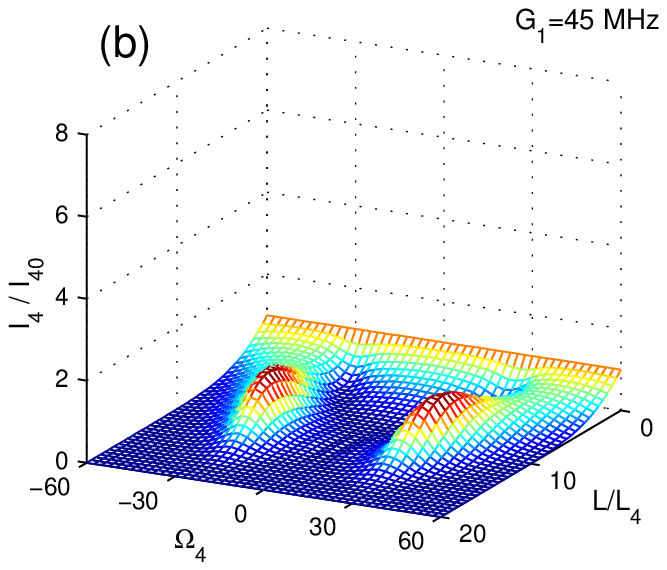}
\includegraphics[width=.32\textwidth]{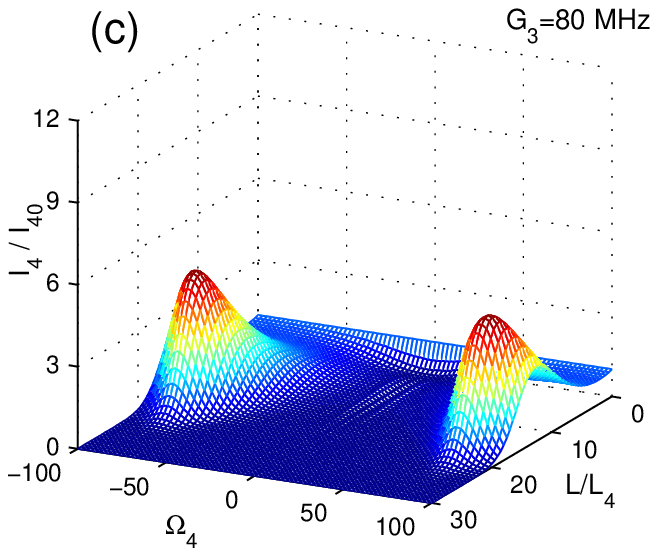}
\end{center}\vspace{-5mm}
\caption{Inversionless gain in optically thick Doppler-broadened medium. $a$ and $c$
-- mov2b.mov (1.50MB, audio$^\ddag$)($G_{10}=60$ MHz, $G_{30}$ varies); $b$ --
mov2a.mov (1.54MB, audio$^\ddag$)($G_{30}=20$ MHz, $G_{10}$ varies). The current
magnitudes of the variable Rabi frequency are displayed at the top of the screen.}
\end{figure}
Figure 3 ($a$-$c$) and the movies mov2a.mov  and mov2b.mov display AWI in an
optically-thick Doppler-broadened medium controlled with injected coherence. Figure
3(a) corresponds to the input intensities, selected for the Fig. 2(a). It shows a
substantial gain, provided by the right choice of the optimum optical length and
resonance detunings. The virtual experiment displays no gain in the resonance. Due to
the interplay of the absorption, Stokes gain and FWM processes as well as depletion
of the driving radiation along the medium, the optimum detuning does not correspond
to that for the maximum Stokes amplification index at the entrance of the cell (Fig.
2(a)). Figure 3(b) is generated at the minimum input driving intensities required to
get the medium transparent. Figure 3(c), which corresponds to the input intensities
used for computing Fig. 2(c), shows that an increase in the intensity of the driving
field $E_3$ even decreases the output signal radiation.
%\begin{floatingfigure}{.25\textwidth}
\vspace {5mm}
\begin{figure}[!h]
\begin{center}
\includegraphics[width=.25\textwidth]{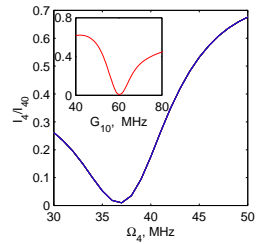}
\vspace{-3mm}\caption{\label{sw} Optical switch.}
\end{center}
\vspace{-5mm}
\end{figure}
%\end{floatingfigure}
It is important to note that a very sharp change in the output
signal at small variation of the frequency or the medium length
(see, e.g., Fig. 3(a)) or the driving intensities (see the movies
mov2a.mov and mov2b.mov) in properly-chosen intervals give rise
to dramatic change in the intensity of the transmitted signal.
This allows one to realize optical switching, based on quantum
coherence processes. As examples, the upper inset in Fig. 4 is
derived from the movies mov2a.mov and mov2b.mov for $G_{30} = 20$
MHz, $\Omega_4 = 37$ MHz and $L /L_{40} = 3$, and the main plot
for $G_{10} = 60$ MHz and $G_{30} = 20$ MHz ($L /L_{40} = 4$). The
minimum transmission is below 1\%. The graphs also show that near
the entrance of the cell, parametric coupling even increases the
rate of depletion of the probe radiation.
\begin{figure}[!h]
\begin{center}
%\hspace*{-5mm}
\includegraphics[width=.24\textwidth]{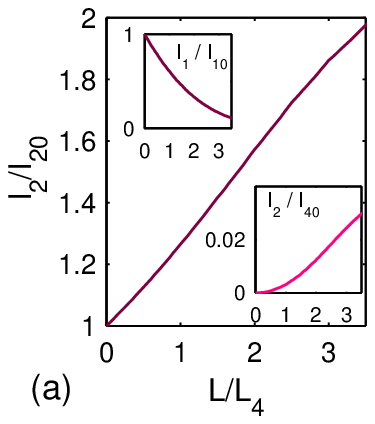}
\includegraphics[width=.24\textwidth]{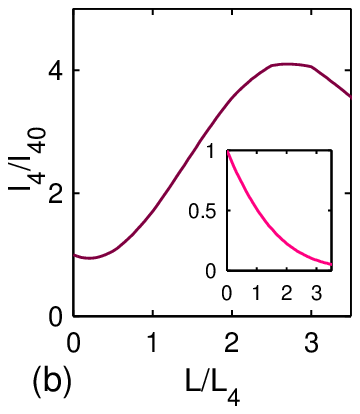}
\includegraphics[width=.24\textwidth]{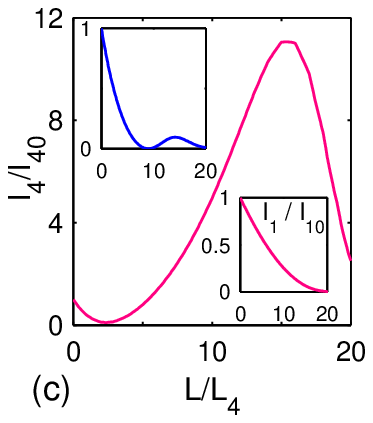}
\includegraphics[width=.24\textwidth]{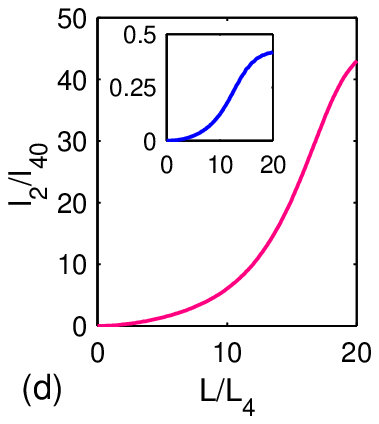}
\end{center}
\vspace{-5mm} \caption{Numerical simulation of the experiment \cite{HIN} -- $(a)$ and
$(b)$,  and of the proposed experiment towards inversionless gain --  $(c)$ and
$(d)$.} \vspace{3mm}
\end{figure}
Figure 5 presents numerical analysis of the experiment \cite{HIN}. In the experiment
all four fields were set in corresponding resonances, the input Rabi frequency at the
Stokes transition being about 10 times larger than the probe one. This was in turn
assumed for computing Figs.5(a) and 5(b). The upper inset in Fig. 5(a) depicts
depletion of the driving radiation $E_1$ ($G_{10}=12$ MHz, $G_{30}=7$ MHz), at the
the optical length, corresponding to the experimental value. The computed gain of the
Stokes beam (main plot in Fig. 5(a)) as well as that for the probe radiation (main
plot in Fig. 5(b)) are in a good agreement with the experimental data. This
simulation is aimed at demonstrating that the gain of the probe radiation is derived
from FWM with no contribution of the parametric amplification. Parametric
amplification in the conditions of this experiment could be achieved by turning off
the input Stokes radiation. Indeed, in this case only depletion of the probe
radiation (lower inset inset in Fig. 5(b)) and weak output Stokes intensity (lower
inset in Fig. 5(a)) come out from the same numerical simulations.

The main plots in the Figs, 5(c,d) are to propose favorable
conditions for inversionless amplification. These plots are
generated for the input driving Rabi frequencies $G_{10}=60$ MHz,
$G_{30}=30$ MHz and for the near optimum detuning $\Omega_4 = 35$
MHz. The main plot in Fig. 5(c) displays substantial inversionless
amplification, and in Fig. 5(d), generated idle radiation under
relatively low input intensities. For the selected transitions a
characteristic value of the Rabi frequency of about 80 MHz
approximately corresponds  to radiation powers in the interval 100
to 200 mW focused on a spot with sizes of about a few parts of a
millimeter. In the same conditions, but for the probe field tuned
to the resonance (at $\Omega_4=0$), gain does not exist and
generated idle radiation dramatically decreases (upper insets in
Figs. 5(c,d)). All the above presented simulations account for phase
mismatch stipulated from the detunings. Both probe and idle
radiations are supposed to be weak and unperturbative.

NIE as an origin of the difference in the rates of induced transitions from upper and
lower levels and consequently -- of AWI were discussed in ref. \cite{NIE}. The
optimum conditions and features of AWI were explored and numerically illustrated for
three level $V$ scheme of neon transitions in ref. \cite{AWI1,AWI2,AWI3}. Related experiments
and evidence of AWI were reported in ref. \cite{BET} (for review of the later works
see ref. \cite{scul1,koch,levi,man,scul2,rau,bull,har}).
In order to simplify the proposed experiment we do not assume
here a source of incoherent excitation of the medium (the formulas accounting for
such excitation are provided in ref. \cite{Mys,akp}). Therefore the energy acquired by
the probe field here is taken entirely from the driving fields. However we call the
effect as inversionless gain, derived from the quantum coherence and interference
processes, rather than optical parametric amplification. This is due to the fact that
the discussed amplification effect can not be thought as the result of the consequent
separate elementary effects of absorption, FWM, Stokes gain and back FWM giving rise
to OPA. We want to stress that consideration of these processes separately in a
resonant medium is not adequate for the physics of the interference process and, as
it has been shown in \cite{TAR}, would result even in qualitatively wrong
predictions. The proposed approach based on NIE increases number of the interfering
channels for the energy conversion including those through populating of the excited
levels, modified Raman-type and parametric processes.

\section{Conclusion}
 In conclusion, this paper considers various optical processes in four-level
Doppler-broa\-dened media, controlled with laser injected coherence. The scheme of
inversionless amplification and optical switching of short-wavelength radiation in
strongly-absorbing resonant four-level gas is proposed. They are controlled with two
driving fields at longer wavelengths and become possible due to the optical
parametric coupling and readily-achievable Stokes gain. Optimal conditions for the
experiment on sodium dimer vapor and the expected gain are explored with the aid of
numerical simulations. The relevant driving intensities correspond to focused cw
radiation on the order of several tens of mW, i.e., to about one photon per thousand
molecules. The analysis shows that further increase of the driving Rabi frequencies
up to about 100-200 MHz allows one to achieve the gain above the threshold of
mirror-less self-oscillation.

AKP thanks B. Wellegehausen and B. Chichkov for stimulating discussions. We thank the
U.S. National Research Council - National Academy of Sciences for support of this
research through the international Collaboration in Basic Science and Engineering
program. AKP and SAM acknowledge funding support from the International Association
of the European Community for the promotion of co-operation with scientists from the
New Independent States of the former Soviet Union (grant INTAS-99-19) and Russian
Foundations for Basic Research  (grant 99-02-39003) and from the Center on
Fundamental Natural Sciences at St. Petersburg University (grant 97-5.2-61).

\end{document}